\documentclass[%
reprint,
superscriptaddress,
amsmath,amssymb,
aps,
pra,
]{revtex4-1}

\usepackage{graphicx}
\usepackage{dcolumn}
\usepackage{bm}
\usepackage{braket}
\usepackage{siunitx}
\usepackage[colorlinks,
linkcolor=blue,
urlcolor=blue,
anchorcolor=blue,
citecolor=blue]{hyperref}
\usepackage{amsmath,amssymb,amsfonts}
\usepackage{epstopdf}
\usepackage[dvipsnames]{xcolor}
\begin{document}
\preprint{APS/123-QED}

\title{Influence of molecular rotation on the generation of N$_2^+$ air lasing} 

\author{Wenli Yang}\affiliation{School of Nuclear Science and Technology, Lanzhou University, Lanzhou 730000, China} 
\affiliation{Frontiers Science Center for Rare Isotopes, Lanzhou University, Lanzhou 730000, China}

\author{Ping Li}\affiliation{School of Nuclear Science and Technology, Lanzhou University, Lanzhou 730000, China} 
\affiliation{Frontiers Science Center for Rare Isotopes, Lanzhou University, Lanzhou 730000, China}

\author{Luzhen Yang}\affiliation{School of Nuclear Science and Technology, Lanzhou University, Lanzhou 730000, China} 
\affiliation{Frontiers Science Center for Rare Isotopes, Lanzhou University, Lanzhou 730000, China}

\author{Jianfeng Guo}\affiliation{School of Nuclear Science and Technology, Lanzhou University, Lanzhou 730000, China} 
\affiliation{Frontiers Science Center for Rare Isotopes, Lanzhou University, Lanzhou 730000, China}

\author{Pengji Ding}\affiliation{School of Nuclear Science and Technology, Lanzhou University, Lanzhou 730000, China} 
\affiliation{Frontiers Science Center for Rare Isotopes, Lanzhou University, Lanzhou 730000, China}

\author{Shan Xue}\email[ ]{xues@lzu.edu.cn}\affiliation{School of Nuclear Science and Technology, Lanzhou University, Lanzhou 730000, China} 
\affiliation{Frontiers Science Center for Rare Isotopes, Lanzhou University, Lanzhou 730000, China}

\author{Hongchuan Du}\email[ ]{duhch@lzu.edu.cn}
\affiliation{School of Nuclear Science and Technology, Lanzhou University, Lanzhou 730000, China} 
\affiliation{Frontiers Science Center for Rare Isotopes, Lanzhou University, Lanzhou 730000, China}

\date{\today}              

\begin{abstract}
	
N$_2^+$ air lasing has attracted considerable attention due to its promising applications in remote sensing and the debates surrounding its generation mechanisms. Here, we present a comprehensive theoretical investigation of the role of molecular rotation in N$_2^+$ lasing at 391 nm ($B^2 \Sigma _u^+(v''=0)\rightarrow X^2 \Sigma _g^+ (v=0)$). By solving the open-system density matrix and Maxwell-Bloch equations in a rovibronic-state basis, we examine both the formation of the N$_2^+$ gain medium induced by a femtosecond pump pulse and the subsequent spatial propagation of the seed pulse. During the pump stage, rotational dynamics are found to significantly modify the angle-dependent populations of ionic vibrational-electronic states within tens of femtoseconds. Furthermore, ionization-produced rotational coherences substantially enhance the population inversion between the $X^2 \Sigma _g^+ (v=0)$ and $B^2 \Sigma _u^+(v''=0)$ states. In the seed propagation stage, both population inversion and rotational coherence are found to contribute to the lasing process, with the latter playing a dominant role in amplifying the lasing signals. These findings reveal the crucial role of molecular rotation in N$_2^+$ air lasing and highlight its potential as a tunable parameter for controlling lasing dynamics.

\end{abstract}
 
\maketitle

\section{Introduction }
When ultrafast intense lasers interact with ambient air, cavity-free directional laser can be generated, known as air lasing. This phenomenon holds great potential for applications in remote sensing and pollutant monitoring  \cite{yao2011high,luo2003lasing,dogariu2011high,hemmer2011standoff,yao2011high,mitryukovskiy2014backward,point2014lasing}\cite{xu2015femtosecond,li2017generation,zhao2020lasing,li2020giant,britton2020control,fu2021extremely}.
Among various air lasing, N$_{2}^{+}$ lasing at 391 nm ($B^2 \Sigma _u^+(v''=0)\rightarrow X^2 \Sigma _g^+ (v=0)$) has attracted significant interest due to ongoing debates regarding its generation mechanism \cite{liu2020extremely,luo2003lasing,yao2011high,wang2013self,ni2014impulsive,xu2015sub,xu2015femtosecond,yao2016population,xu2017alignment,liu2017unexpected,zhong2017vibrational,arissian2018transient,zhang2019coherent,li2019significant,kleine2022electronic,mysyrowicz2019lasing,richter2020rotational,lytova2020n} and its rich exhibition of strong-field physical phenomena. To date, several theoretical models have been proposed to explain the generation mechanism of the N$_{2}^{+}$ 391-nm lasing. These models are built upon two types of mechanisms: (1) the population inversion mechanism \cite{xu2015femtosecond,yao2016population,li2019significant,kleine2022electronic}, and (2) the lasing without inversion (LWI) mechanism  \cite{mysyrowicz2019lasing,richter2020rotational,lytova2020n}. Under seed pulse excitation, both mechanisms can lead to the buildup of macroscopic polarization in the N$_2^+$ ensemble, ultimately resulting in lasing emission.
 
An important factor influencing the generation of N$_2^+$ lasing is the effect of molecular rotation. In fact, pump-probe experiments have demonstrated that molecular rotation can induce periodic oscillations in the delay-dependent lasing signals on a picosecond timescale \cite{zhang2013rotational,xie2014coupling,ando2019rotational}. Furthermore, the temporal profile of the lasing electric field also exhibits picosecond rotational periods \cite{li2014signature,xie2021controlling}.   
Driven by these experimental observations, theoretical studies on the role of molecular rotation have progressively increased in recent years. Specifically, its effects have been examined in two distinct stages: the preparation of the N$_{2}^{+}$ ensemble through pump laser excitation and the subsequent amplification of the seed pulse. 
For the pump stage, Zhang \textit{et al.}~examined the establishment of population inversion between the rotational states within the $X^2 \Sigma _g^+ $ and $B^2 \Sigma _u^+$ states by solving the Schr$\ddot{\text{o}}$dinger equation. They found that population inversion can occur between specific rotational states even in the absence of inversion between electronic states \cite{zhang2020rotationally}. 
They further investigated the impact of ionization timing on the post-ionization dynamics of N$_2^+$. Their results indicated that when the laser intensity exceeds $4.0 \times 10^{14}$W/cm$^2$, ionization events must be considered throughout the entire duration of the laser pulse \cite{zhang2022effects}. 
Recently, we found that not only the ionization-produced ionic populations, but also the ionic coherences, can influence the post-ionization dynamics within N$_2^+$. The later leads to a noticeable increase in the population inversion between the $X^2 \Sigma _g^+ (v=0)$ and $B^2 \Sigma _u^+(v''=0)$ states. However, the role of coherences among rotational-vibrational-electronic (rovibronic) states produced by strong-field ionization has not yet been fully explored. How these coherences affect the molecular dynamics of the residual ions remains an open question.
For the seed amplification stage, Richter \textit{et al.} demonstrated that rotational coherences (RCs), excited by the pump pulse, can modulate the gain and absorption of the delayed seed pulse \cite{richter2020rotational}. These coherences can induce $B^2 \Sigma _u^+(v''=0) \rightarrow X^2 \Sigma _g^+(v=0)$ optical emission, even in the absence of population inversion \cite{richter2020rotational,lytova2020n}. This effect is attributed to the lasing without inversion (LWI) mechanism, which involves quantum interference between different rotational transition pathways. 
Despite these findings, several open questions remain. To what extent do RCs enhance the lasing signal during spatial propagation through the N$_2^+$ gain medium? Can RCs and population inversion jointly contribute to seed amplification? These aspects require further investigation.

In this study, we solve the open-system density matrix (DM) and the Maxwell-Bloch (MB) equations to investigate the effects of molecular rotation on N$_2^+$ 391-nm lasing. First, we examine the pump process with explicit consideration of the ionization-produced RCs. Our simulations reveal that these coherences drive ultrafast population redistribution among rovibronic states within just tens of femtoseconds, highlighting the importance of rotational dynamics in establishing population inversion. Subsequently, we examine the seed amplification process and demonstrate that the combined effects of population inversion and RC contribute to the generation of the N$_{2}^{+}$ lasing, with the latter playing a dominant role in the seed amplification. 

This paper is organized as follows. Section II describes the theoretical methods. Sections III~A and III~B examine the role of molecular rotation in the pump stage, while Sec.~III~C focuses on its influence in the seed propagation stage. Section IV concludes the paper. Atomic units (a.u.) are used throughout, unless otherwise specified.

\section{THEORETICAL METHOD}
\subsection{Preliminaries}
\textbf{Basis set:} When an intense laser pulse interacts with N$_2$, strong-field ionization primarily occurs from four molecular orbitals (MO): $3\sigma_g$ [highest occupied MO (HOMO)], $1\pi_{u\mp}$ (degenerate HOMO-1), and $2\sigma_u$ (HOMO-2), resulting in the formation of N$_2^+$ in the $X^2 \Sigma _g^+ $, $A^2 \Pi _{u\pm}$, and $B^2 \Sigma _u^+$ electronic states, respectively. For conciseness, these states are abbreviated as $X$, $A\pm$, and $B$ hereafter. For convenience, $A$ represents the combination of the $A+$ and $A-$ states. 
 
To incorporate the effects of molecular rotation, we adopt the rovibronic states $|ivJMK \rangle$ as the basis set. Here, $i=X$, $A\pm$, or $B$ represents the electronic state in N$_2^+$. $v$ denotes the level of the vibrational state. $J$ is the total angular momentum quantum number. $M$ and $K$ are the magnetic quantum numbers with respect to the space-fixed (SF) $Z$ axis and the molecule-fixed (MF) $z$ axis (molecular axis), respectively. 
Note that $K$ depends on the electronic state. For the $X$ and $B$ states, $K=0$, while for the $A\pm$ states, $K=\pm 1$.
Under the Born-Oppenheimer approximation, the rovibronic-state wavefunction can be expressed as a product of the wavefunctions of the electronic state $\psi_i$, the vibrational state $\chi^i_v$, and the rotational state $\Theta_{JMK}$, 
\begin{equation} \label{basis}
|ivJMK \rangle=\psi_i(\mathbf{r};R) \chi^i_v(R) \Theta_{JMK}(\alpha, \beta, \gamma),  
\end{equation}
where $\mathbf{r}$ and $R$ represent the electronic coordinates and the internuclear distance, respectively. $\alpha$, $\beta$, and $\gamma$ denote the Euler angles in the $ZYZ$ convention, which define the orientation of the molecular axis.
The rotational state $\Theta_{JMK}$ is represented by the Wigner $D$-function as
\begin{equation} \label{dfunc}
\begin{aligned} 
\Theta_{JMK} (\alpha, \beta, \gamma )&=\sqrt{\frac{2J+1}{8\pi^2} }{D^J_{MK}}^*(\alpha \beta \gamma)\\
&=\sqrt{\frac{2J+1}{8\pi^2} }e^{iM\alpha }{d^J_{MK}}^*(\beta)e^{iK\gamma }.
\end{aligned}
\end{equation}%
The eigenenergy of the rovibronic state is given by
\begin{equation} \label{energy}
\begin{aligned}
E^i_{vJ}&=E^i_v+\left [ B_i-\alpha_i( v+\frac{1}{2}) \right ] J(J+1)-D_iJ^2(J+1)^2.
\end{aligned}
\end{equation}%
Here, $E^i_v$ denotes the energy of the vibrational-electronic (vibronic) state. $B_i$ is the rotational constant of the $i$-th electronic state. $\alpha^i$ and $D^i_e$ are the anharmonicity correction and centrifugal distortion constant, respectively. These constants are taken from the NIST Chemistry Webbook \cite{nist2025}.  


\textbf{Molecule-laser interaction:}
Under the dipole approximation and in the length gauge, the molecule-laser interaction operator is defined as $\hat{O}=-\mathbf{r}^{\text{(MF)}}\cdot \mathbf{F}^{\text{(MF)}}$, where $\mathbf{r}^{\text{(MF)}}$ and $\mathbf{F}^{\text{(MF)}}$ represent the electronic coordinate and the electric field in the MF $xyz$ coordinate system, respectively. $\mathbf{F}^{\text{(MF)}}$ can be transformed from $\mathbf{F}^{\text{(SF)}}$, which is defined in the SF $XYZ$ coordinate system. In this work, both the pump and seed pulses are polarized along the SF $Z$ axis. The electric field can be expressed as   
\begin{equation} \label{field}
\begin{aligned} 
&\mathbf{F}^{\text{(SF)}}=F_Z\mathbf{e}_Z\\
&\mathbf{F}^{\text{(MF)}}=-F_Z\text{sin}\beta\text{cos}\gamma \mathbf{e}_x+F_Z\text{sin}\beta\text{sin}\gamma\mathbf{e}_y+F_Z\text{cos}\beta\mathbf{e}_z.
\end{aligned}  
\end{equation}%
Then, the interaction term in the basis of rovibronic states is given by
\begin{equation} \label{eq:oij1}
\begin{aligned} 
O^{ij}_{vJMK,v'J'M'K'}&=\left \langle ivJMK |\hat{O}|jv'J'M'K' \right \rangle \\
&=\left \langle vJMK |\mathbf{u}_{ij}(R)\cdot \mathbf{F}^{\text{(MF)}}|v'J'M'K' \right \rangle\\
&=\left \langle JMK |\mathbf{u}_{vv'}^{ij}\cdot \mathbf{F}^{\text{(MF)}}|J'M'K' \right \rangle.\\
\end{aligned} 
\end{equation}%
Here, $\mathbf{u}_{ij}(R)$ is the $R$-dependent electronic-state transition dipole moment (TDM) between electronic states $i$ and $j$, and $\mathbf{u}_{vv'}^{ij}=\left \langle \chi ^i_v(R)|\mathbf{u}_{ij}(R)|\chi ^j_{v'}(R) \right \rangle $ is the vibronic-state TDM.
In N$_2^+$, there exist $X-B$ and $X-A\mp$ TDMs, which are oriented parallel and perpendicular to the molecular axis, respectively. Specifically, these two TDMs can be expressed as $\mathbf{u}_{XA\mp}=(\mp u_{\perp} \mathbf{e}_x+\text{i}u_{\perp}\mathbf{e}_y)/\sqrt{2}$ and $\mathbf{u}_{XB}=u_\parallel$, respectively \cite{PhysRevA.108.023123, xue2025}. Following this convention, Eq.~(\ref{eq:oij1}) can be further written as
\begin{equation} \label{eq:oij2}
\begin{small}
\begin{aligned} 
O^{XB}_{vJM0,v'J'M'0}&=\left \langle JM0|u^\parallel_{vv'}  \text{cos}\beta |J'M'0\right \rangle F_Z\\
&=u^{XB}_{vJM0,v'J'M'0}F_Z\\
O^{XA_\pm}_{vJM0,v'J'M'\pm1}&=\left \langle JM0|\mp\frac{u^\perp_{vv'} }{\sqrt{2}} \text{sin}\beta e^{\mp i\gamma}|J'M'\pm1\right \rangle F_Z\\
&=u^{XA_\pm}_{vJM0,v'J'M'\pm1}F_Z,\\
\end{aligned} 
\end{small}
\end{equation}%
where $u^\parallel_{vv'} =\left \langle \chi ^i_v |u_\parallel |\chi ^j_{v'} \right \rangle$ and $u^\perp_{vv'}=\left \langle \chi ^i_v |u_\perp |\chi ^j_{v'}  \right \rangle$.
The quantities $u^{XB}_{vJM0,v'J'M'0}$ and $u^{XA_\pm}_{vJM0,v'J'M'\pm1}$ represent the rovibronic-state TDMs.
Since the diatomic molecule exhibits cylindrical symmetry with respect to the molecular axis, $\gamma$ is set to $0^\circ$ for convenience. 
Using the properties of the Wigner $D$-function, the rovibronic-state TDMs can be expressed as
\begin{equation} \label{eq:oij3}
\begin{small}
\begin{aligned} 
&u^{XB}_{vJM0,v'J'M'0}\\
=&u^\parallel_{vv'} \times   
\left\{\begin{matrix}
 \sqrt{  \frac{(J-M)(J+M)}{(2J-1)(2J+1)}  } & J'=J-1,M=M'\\
 \sqrt{  \frac{(J-M+1)(J+M+1)}{(2J+1)(2J+3)}  }  & J'=J+1,M=M'
\end{matrix}\right.
\\
&u^{XA_\pm}_{vJM0,v'J'M'\pm1}\\
=&u^\perp_{vv'}\times  
\left\{\begin{matrix}
 \pm\sqrt{  \frac{M}{2J(2J+1)}  } & J'=J,M=M'\\
 \sqrt{  \frac{(J-M)(J+M)(J-1)}{2J(2J-1)(2J+1)}  }  & J'=J-1,M=M'\\
 -\sqrt{  \frac{(J-M+1)(J+M+1)(J+2)}{2(J+1)(2J+1)(2J+3)}  }  & J'=J+1,M=M'.
\end{matrix}\right.
\end{aligned} 
\end{small}
\end{equation}%

\textbf{Thermal distribution:}
Prior to the arrival of the laser pulse, N$_2$ is assumed to be prepared in an incoherent ensemble. The rovibronic states are distributed according to the Boltzmann distribution. At room temperature ($T \approx  300~K$), the populations of excited electronic and higher vibrational states are negligible, and the neutral molecules primarily occupy the vibronic ground state. Consequently, the molecular ensemble can be effectively described by a Boltzmann distribution over the rotational states $\left |  J_0M_0 \right \rangle $ in the vibronic ground state 
\begin{equation} \label{thermal}
C_{J_0}^{Bol}=C_Ng_{J_0}e^{-E_{J_0}/k_BT},
\end{equation}
where $C_N$ is the normalization constant. $g_{J_0}$ is the statistical weight of the nuclear spin, which takes the values of $g_{J_0}=2$ for even $J_0$ and $g_{J_0}=1$ for odd $J_0$. $E_{J_0}$ is the eigenenergy of state $\left |  J_0M_0 \right \rangle $, and $k_B$ is the Boltzmann constant.

\subsection{Quantum Liouville equations and coherence models}
With the basis set and interaction matrix elements defined, the ionic DM $\rho(t)$ can be propagated in time using the quantum Liouville equations
\begin{equation} \label{eq:QLE}
\frac{\partial}{\partial t}\rho(t)= -\text{i}[H(t),\rho(t)]+f(\rho(t)),
\end{equation}%
where $H=H_0+\mathbf{r}\cdot\mathbf{F}(t)$ denotes the Hamiltonian of the N$_2^+$ system. $f(\rho(t))$ provides a phenomenological description of the ionization (N$_2\rightarrow$N$_2^+$) and decoherence processes.  

\textbf{Vibronic-state coherence model:} In the absence of molecular rotation, the DM elements are expressed in the vibronic-state basis as $\rho_{vv'}^{ij}=\langle iv |\hat{\rho}|jv' \rangle$, where $\hat{\rho}$ is the density operator. Substituting $\rho_{vv'}^{ij}$ into Eq.~(\ref{eq:QLE}) produces a set of coupled differential equations
\begin{equation} \label{eq:QLEv}
\begin{aligned}
\frac{\partial}{\partial t}\rho_{vv'}^{ij}&=-\text{i}\omega_{vv'}^{ij}\rho_{vv'}^{ij}\\
&\text{i}\mathbf{F}^{\text{(MF)}}\cdot\sum_{k,v''}(\mathbf{u}_{vv''}^{ik}\rho_{v''v'}^{kj}-\rho_{vv''}^{ik}\mathbf{u}_{v''v'}^{kj})\\
&-\rho_{vv'}^{ij}/T_d+\rho_{vv'}^{ij\text{(ins)}}, 
\end{aligned}
\end{equation}%
where $\omega_{vv'}^{ij}=E_{v}^{i}-E_{v'}^{j}$ is the energy difference between the vibronic states. $T_d$ is the decoherence time for $\rho_{vv'}^{ij}$ with $i\ne j $ or $v\ne v'$, and is set to 1~ps during the pump stage \cite{lei2022ultraviolet}.
$\rho_{vv'}^{ij\text{(ins)}}(t)$ represents the instantaneous reduced DM (RDM) produced by strong-field ionization at each ionization instant \cite{Pabst2016,zhu2023influence}. The diagonal terms represent the ionization rates from the neutral ground state to different ionic vibronic states, while the off-diagonal terms describe the ionization-produced coherences. Based on the coherence model developed in Refs.~\cite{Pabst2016,xue2021vibronic,xue2025}, $\rho_{vv'}^{ij\text{(ins)}}(t)$ can be expressed as
\begin{equation} \label{eq:insv}
\begin{aligned}
&\rho_{vv'}^{ij\text{(ins)}}(t)=n(t)\times\langle iv|0 \rangle \langle 0|jv' \rangle \\
&\times\sqrt{\Gamma_i(\beta)\Gamma_j(\beta)}\prod_{\xi =x,y,z}\text{sgn}[F_{\xi}^{\text{(MF)}}(t)]^{(2-P_i^\xi-P_j^\xi)/2}.\\
\end{aligned}
\end{equation}%
Here, $\langle iv|0 \rangle$ is the overlap integral between the wavefunctions of the neutral vibronic ground state and the ionic vibronic state. 
$\Gamma_i(\beta)$ represents the ionization rate to the $i$th ionic state, which is calculated using the molecular Ammosov-Delone-Krainov (MO-ADK) theory. $\beta$ is the angle between the molecular axis and the SF $Z$ direction. 
$\text{sgn}(x)$ is the sign function. $P_{i(j)}^\xi$ describes the reflection parity of the $i(j)$th ionizing orbital with respect to the plane perpendicular to the MF $\xi$ axis, with a value of $\pm 1$ for $\pm$ parity. $n(t)$ is the remaining probability of the neutral molecule at time $t$.

To note, rigorously calculating the ionization-produced coherence requires tracing out the free-electron degrees of freedom, as is approximately done in the partial-wave expansion method or the adiabatic strong-field approximation method \cite{PhysRevA.108.023123,xue2025}. Since Eq.~(\ref{eq:insv}) does not incorporate this procedure, the coherence is overestimated. In fact, according to Eq.~(\ref{eq:insv}), the degree of coherence (DOC) between electronic states is 100\%, where the DOC is defined as $|\rho_{ij}|/\sqrt{\rho_{ii}^2\rho_{jj}^2}$ with $i$ and $j$ denoting the state levels. Detailed discussions of different coherence models can be found in Ref.~\cite{xue2025}.

\textbf{Rovibronic-state coherence model:} To incorporate molecular rotation, we use rovibronic states as the basis set, and express the DM element as $\rho_{vJMK,v'J'M'K'}^{ij}=\langle ivJMK |\hat{\rho}|jv'J'M'K' \rangle$. Substituting it into Eq.~(\ref{eq:QLE}), one obtains
\begin{widetext}  
\begin{equation} \label{QLEr} 
\begin{aligned}
\frac{\partial}{\partial t}\rho_{vJMK,v'J'M'K'}^{ij}&=-i\omega^{ij}_{vJ,v'J'}\rho_{iJMK,jJ'M'K'}^{mn}\\
&+i\sum_{kv''J''\!M''K''} \!\!(O_{vJMK,v''J''M''K''}^{ik}\rho_{v''J''\!M''\!K'',v'J'M'K'}^{kj}
-\rho_{vJMK,v''J''\!M''\!K''}^{ik}  O_{v''J''\!M''\!K'',v'J'M'K'}^{kj}   )\\
&- \rho^{ij}_{vJMK,v'J'M'K'}/T_d  + \rho^{ij\text{(ins)}}_{vJMK,v'J'M'K'},	
\end{aligned}
\end{equation}
\end{widetext}
where $\omega^{ij}_{vJ,v'J'}=E^i_{vJ}-E^j_{v'J'}$ is the energy difference between rovibronic states.  
$\rho^{ij(\text{ins})}_{vJMK,v'J'M'K'}$ represents the instantaneous ionization-produced RDM of N$_2^+$. To derive its expression, we assume that the orientation of N$_2^+$ is predominantly determined by the geometric alignment induced by strong-field ionization \cite{posthumus1998dynamic}, while the influence of the rotational evolution of N$_2$ is neglected. The latter has only a minor effect on the lasing signal, as discussed in Ref.~\cite{lytova2020n}.
As an extension of Eq.~(\ref{eq:insv}), $\rho^{ij(\text{ins})}_{vJMK,v'J'M'K'}$ is expressed as
\begin{widetext}
\begin{equation} \label{eq:insr}
\begin{aligned}
\rho^{ij(\text{ins})}_{vJMK,v'J'M'K'}&= \langle iv|0 \rangle \langle 0|jv' \rangle \\
&\times\sum_{J_0M_0}  C^{Bol}_{J_0}\left\langle JMK\left|\sqrt{\Gamma_i(\beta) }\prod_{\xi =x,y,z}\text{sgn}[F_{\xi}^{\text{(MF)}}(t)]^{(1-P_i^\xi)/2}\right | J_0M_0^{(\text{r})}\right\rangle \\ 
&\times \left\langle J_0M_0^{(\text{r})}\left |\sqrt{\Gamma_j(\beta) }\prod_{\xi =x,y,z}\text{sgn}[F_{\xi}^{\text{(MF)}}(t)]^{(1-P_j^\xi)/2}\right|J'M'K'\right\rangle. \\
\end{aligned}     
\end{equation}
\end{widetext} 
Here, $|J_0M_0^{(\text{r})}(\alpha,\beta,t)\rangle$ represents the time-dependent rotational wavefunction during laser interaction, which is initially prepared in the $|J_0M_0\rangle$ state and satisfies
\begin{equation} \label{eq:insr2}
\begin{aligned}
&\frac{d}{dt}\left||J_0M_0^{(\text{r})}(\alpha,\beta,t)\rangle\right|^2\\
=&-\left(\sum_i \Gamma_i(\beta)\right) \left||J_0M_0^{(\text{r})}(\alpha,\beta,t)\rangle\right|^2.
\end{aligned}
\end{equation}
$|JMK\rangle$ and $|J'M'K'\rangle$ represent the ionic rotational states associated with the vibronic states $|iv\rangle$ and $|jv'\rangle$, respectively. 
Note that in the cylindrically symmetric system, where the laser electric fields are polarized along the SF $Z$ axis, the magnetic quantum number along the SF $Z$ axis remains unchanged. Therefore, $\rho^{ij(\text{ins})}_{vJMK,v'J'M'K'}$ is non-zero only when $M_0=M=M'$. 

To compare with the vibronic-state model, we calculate the angle-dependent population of the vibronic states as follows
\begin{equation}\label{vibrodis}
\rho_{vv}^{ii}(\beta )=\sum_{J,J',M}  \rho_{vJMK,vJ'MK}^{ii}(t_{\text{end}})\Theta_{JMK}\Theta_{J'MK}^*,
\end{equation}
where $t_{\text{end}}$ represents the end time of the pump pulse.


\subsection{Maxwell-Bloch equations}
Once the N$_2^+$ gain medium is prepared by the pump pulse, a delayed seed pulse can be employed to stimulate the lasing emission. The propagation of the seed pulse is modeled by solving the MB equations \cite{zhu2023influence}. 
Considering that both the seed and pump pulses are polarized along the SF $Z$ direction, the stimulated lasing field is also polarized along the same direction \cite{xie2020role}. 
For the 391-nm N$_2^+$ lasing, the electric field as functions of time and propagation direction $Y$ is expressed as
\begin{small}
\begin{equation} \label{lasing}
\begin{aligned}
&F^{\text{lasing}}(t,Y)\\
=& \frac{1}{2}\left(\!\tilde {F}^{\text{seed}}(t,Y)e^{-i\omega_s t}\!+\!\sum_{J'JM}\tilde {F}^{BX}_{J'JM}(t,Y)e^{-i\omega^{BX}_{0J',0J} t}\!\right)+c.c.\\ 
=&\frac{1}{2}\mathcal {F}^{\text{lasing}}(t,Y) +c.c.,
\end{aligned}
\end{equation}
\end{small}%
where $\tilde {F}^{\text{seed}}(t,Y)$ is the slowly varying envelope of the seed pulse centered at frequency $\omega_s$, and $\tilde {F}^{BX}_{J'JM}(t,Y)$ represents the envelope of the stimulated lasing field at frequency $\omega^{BX}_{0J',0J}$. The stimulated field is zero at $Y = 0$.
Under the rotational wave approximation, the simplified Maxwell equation in the retarded frame ($t'=t-Y/c$) can be expressed as
\begin{small}
\begin{equation} \label{eq:MAXWELL}
\begin{aligned}
\frac{\partial }{\partial Y}\tilde {F}^{BX}_{J'JM}(t',Y)&=i\alpha^{BX}_{0J',0J}\tilde{\rho}^{BX}_{0J'M,0JM}(t',Y)\\
\end{aligned}
\end{equation}
\end{small}%
Here, $\alpha^{BX}_{0J',0J}=N\frac{\omega^{BX}_{0J',0J}u^{BX}_{0J'M,0JM}}{\varepsilon_0c}  $ represents the propagation parameter, where $\varepsilon_0$ is the vacuum permittivity, $c$ is the speed of light, and $N$ represents the density of N$_2^+$. 
$\tilde{\rho}^{BX}_{0J'M,0JM}=\rho^{BX}_{0J'M,0JM} e^{i \omega^{BX}_{0J',0J} t'}$ is the slow-varying rovibronic-state coherence.

To calculate the coherence, we solve the multi-state Bloch equations that describe the interaction of the rovibronic states $X_{0,JM}$ and $B_{0,J'M'}$ with the lasing field, which are given by
\begin{widetext}
\begin{small}
\begin{equation} \label{eq:BLOCH}
	\begin{aligned}
&\frac{\partial }{\partial t}\tilde{\rho}_{0JM,0J'M}^{XB}
=i\frac{1}{2}\mathcal {F}^{\text{lasing}*}e^{-i\omega^{BX}_{0J^{\prime},0J}t}
\left(\sum_{J''}u_{0JM,0J''M}^{XB}\rho^{BB}_{0J''M,0J'M}
-\sum_{J''}\rho^{XX}_{0JM,0J''M}u_{0J''M,0J'M}^{XB}  \right)-\frac{\tilde{\rho}_{0JM,0J'M}^{XB}}{T_d},\\
&\frac{\partial }{\partial t}{\rho}_{0JM,0J'M}^{XX}
=-i\omega^{XX}_{0JM,0J'M}\rho_{0JM,0J'M}^{XX} +
\frac{1}{2}\left(\sum_{J''} u_{0JM,0J''M}^{XB} \rho^{BX}_{0J''M,0J'M}\mathcal {F}^{\text{lasing}*}
-\sum_{J''}\rho^{XB}_{0JM,0J''M}u_{0J''M,0J^{\prime}M}^{BX}\mathcal{F}^{\text{lasing}}\right)-\frac{\rho_{0JM,0J'M}^{XX}}{T_d}, \\
&\frac{\partial }{\partial t}{\rho}_{0JM,0J'M}^{BB} 
=-i\omega^{BB}_{0JM,0J'M}\rho_{0JM,0J'M}^{BB} +
\frac{1}{2}\left(\sum_{J''} u_{0JM,0J''M}^{BX} \rho^{XB}_{0J''M,0J^{\prime}M}\mathcal{F}^{\text{lasing}}
-\sum_{J''}\rho^{BX}_{0JM,0J''M}u_{0J''M,0J'M}^{XB}\mathcal{F}^{\text{lasing}*}\right)-\frac{\rho^{BB}_{0JM,0J'M}}{T_d},
		\end{aligned}%
		\end{equation}
\end{small}%
\end{widetext}
where $T_d$ is the collision-induced decoherence time, and is set to 5~ps for the seed propagation stage \cite{zhu2023influence}.
By solving the coupled Eqs.~(\ref{eq:MAXWELL}) and (\ref{eq:BLOCH}), the propagation of the seed pulse can be determined, which ultimately leads to the generation of the lasing electric field $F^{\text{lasing}}$.

\section{Results and Discussion}
\subsection{Comparison between vibronic and rovibronic models}

\begin{figure} [htb]
\includegraphics[width=8.5cm,clip=true]{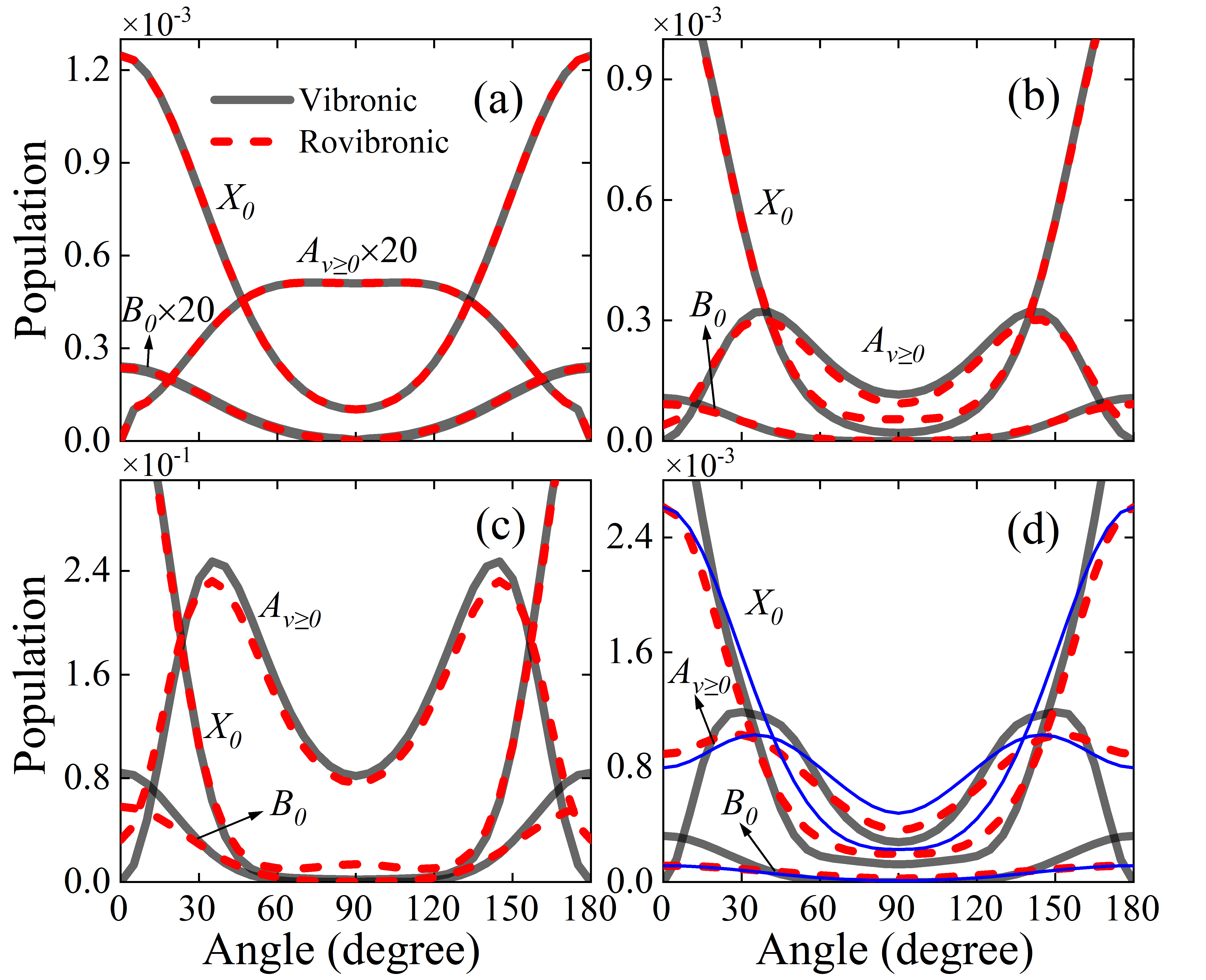}
\caption{Angle-dependent populations of the $X_{0}$, $A_{v'\ge 0}$, and $B_{0}$ states calculated by the vibronic and rovibronic models.
(a) Considering only the ionization process. For clarity, data for $A_{0}$ and $B_{0}$ are magnified by 20 times.
(b) presents the calculation results of the vibrational and rotational model under laser parameters $\tau=10\ \mathrm{fs}, I=1\times 10^{14}\ \mathrm{W/cm^{2}}, \lambda=800\ \mathrm{nm}$.
(c) shows the calculation results of the vibrational and rotational model under laser parameters $\tau=10\ \mathrm{fs}, I=3\times 10^{14}\ \mathrm{W/cm^{2}}, \lambda=800\ \mathrm{nm}$.
(d) presents the calculation results of the vibrational and rotational model under laser parameters $\tau=30\ \mathrm{fs}, I=1\times 10^{14}\ \mathrm{W/cm^{2}}, \lambda=800\ \mathrm{nm}$. The data represented by the blue thin lines are obtained by applying 45 fs of free evolution to the calculation results with laser parameters $\tau=10\ \mathrm{fs}, I=1\times 10^{14}\ \mathrm{W/cm^{2}}, \lambda=800\ \mathrm{nm}$, and normalized to the $X_{0}$, $A_{0}$, and $B_{0}$ values of the red dashed lines.}
\label{fig1}
\end{figure}

In this section, we compare the vibronic and rovibronic models to examine how molecular rotation affects the N$_2^+$ vibronic-state populations after strong-field ionizations of the neutral ensemble.
Figure~\ref{fig1} shows the angle-dependent populations of the $X_0$, $B_0$ and $A_{v'\ge 0}$ states at the end of the pump pulse, calculated using both the vibronic and rovibronic models. For the rovibronic model, the angle-dependent vibronic-state populations are calculated using Eq.~(\ref{vibrodis}). These states are examined because they are closely related to N$_2^+$ 391-nm lasing. The population inversion between the $X_0$ and $B_0$ states is widely recognized as a necessary condition for generating the 391-nm lasing. The $A_{v'\ge 0}$ states are considered to act as a ``reservoir" to deplete the $X_0$-state population \cite{yao2016population,zhu2023influence}.

First, we examine the angle-dependent populations resulting solely from strong-field ionizations for the two models. An 800-nm, 10-fs (FWHM) pump pulse with a peak intensity of $1 \times10^{14}$W/cm$^{2}$ is applied. Dipole coupling and the free evolution of rotational states are excluded. As shown in Fig.~\ref{fig1}(a), the vibronic and rovibronic models produce identical results. This indicates that the ionization dynamics are treated consistently for the two models. 
Next, we include the effects of dipole coupling and the rotational-state evolution. Corresponding angle-dependent populations are shown in Fig.~\ref{fig1}(b). The populations of the $B_0$ and $A_{v'\ge 0}$ states exhibit notable enhancement. Nevertheless, the results from both models remain nearly identical. We then increase the peak laser intensity to $3 \times10^{14}$W/cm$^{2}$. Corresponding results are shown in Fig.~\ref{fig1}(c). At this higher intensity, the angle-dependent populations predicted by the rovibronic model begin to diverge from those obtained using the vibronic model. These discrepancies primarily arise from the enhanced RCs, $ \rho_{vJMK,vJ'MK}^{ii}$, which are induced by stronger transitions between rovibronic states under the more intense laser field. As indicated by Eq.~(\ref{vibrodis}), these coherences affect the angle-dependent vibronic-state populations.

Finally, we examine the effect of pulse duration. Keeping the peak laser intensity the same as in Fig.~\ref{fig1}(b), we increase the pulse duration to 30~fs. Corresponding results are shown in Fig.~\ref{fig1}(d).   
In contrast to Fig.~\ref{fig1}(b), significant discrepancies emerge between the two models.
Since only the pulse durations are different, we attribute these discrepancies to the free evolutions of rotational states. 
To verify this, we assume that the ionic states are predominantly populated at the peak of the laser envelope. For the 30-fs and 10-fs (FWHM) pulses, the rotational states evolve for approximately 66~fs and 22~fs until the end of the laser pulses, respectively. 
Based on this assumption, we use the rovibronic-state DM calculated at the end of the 10-fs laser pulse as an initial DM, and propagate it freely for an additional 44~fs (66$-$22~fs). 
The freely evolved ``populations" are then calculated using Eq.~(\ref{vibrodis}) and are shown as blue thin lines in Fig.~\ref{fig1}(d).  
The results exhibit similar trends to those obtained from the rovibronic model, confirming that the discrepancies between the two models in Fig.~\ref{fig1}(d) primarily arises from the free evolution of rotational states. 
Essentially, strong-field ionization of N$_2$ will induce alignment of N$_2^+$ along the field polarization through the so-called geometric alignment \cite{posthumus1998dynamic}. However, due to the rapid dephasing of rotational states, this alignment decays quickly within tens to hundreds of femtoseconds, resulting in significant changes in the rotational wavepacket and, consequently, in the angle-dependent vibronic-state populations. 
These results highlight that, although the rotational period of N$_2^+$ is on the picosecond timescale, molecular rotation cannot be neglected in the post-ionization dynamics of N$_2^+$, even on a tens-of-femtosecond timescale.

\subsection{Effect of ionization-produced coherences}
Strong-field ionization from multiple orbitals not only populates different ionic states but also generates coherences among these states \cite{xue2025}. In this section, we use the rovibronic model to investigate the role of ionization-produced coherences in the transition dynamics of N$_2^+$.
Three types of coherences will be examined:\\
(i) RC, denoted by $\rho^{ii(\text{ins})}_{vJMK,vJ'MK}$, where $J\ne J'$.\\
(ii) Vibrational coherence (VC), denoted by $\rho^{ii(\text{ins})}_{vJMK,v'J'MK}$, where $v \ne v'$.\\
(iii) Electronic coherence (EC), denoted by $\rho^{ij(\text{ins})}_{vJMK,v'J'MK'}$, where $i \ne j$ and $K(K')$ is associated with $i(j)$.\\ 
In the simulations, we incrementally incorporate the RCs, VCs, and ECs to isolate the effects of these three types of coherences.
A linearly polarized 800-nm, 30-fs laser pulse with an intensity of $3\times 10^{14}\text{W/cm}^2$ is employed as the pump pulse.


\begin{figure} [htb]
\includegraphics[width=9.5cm,clip=true]{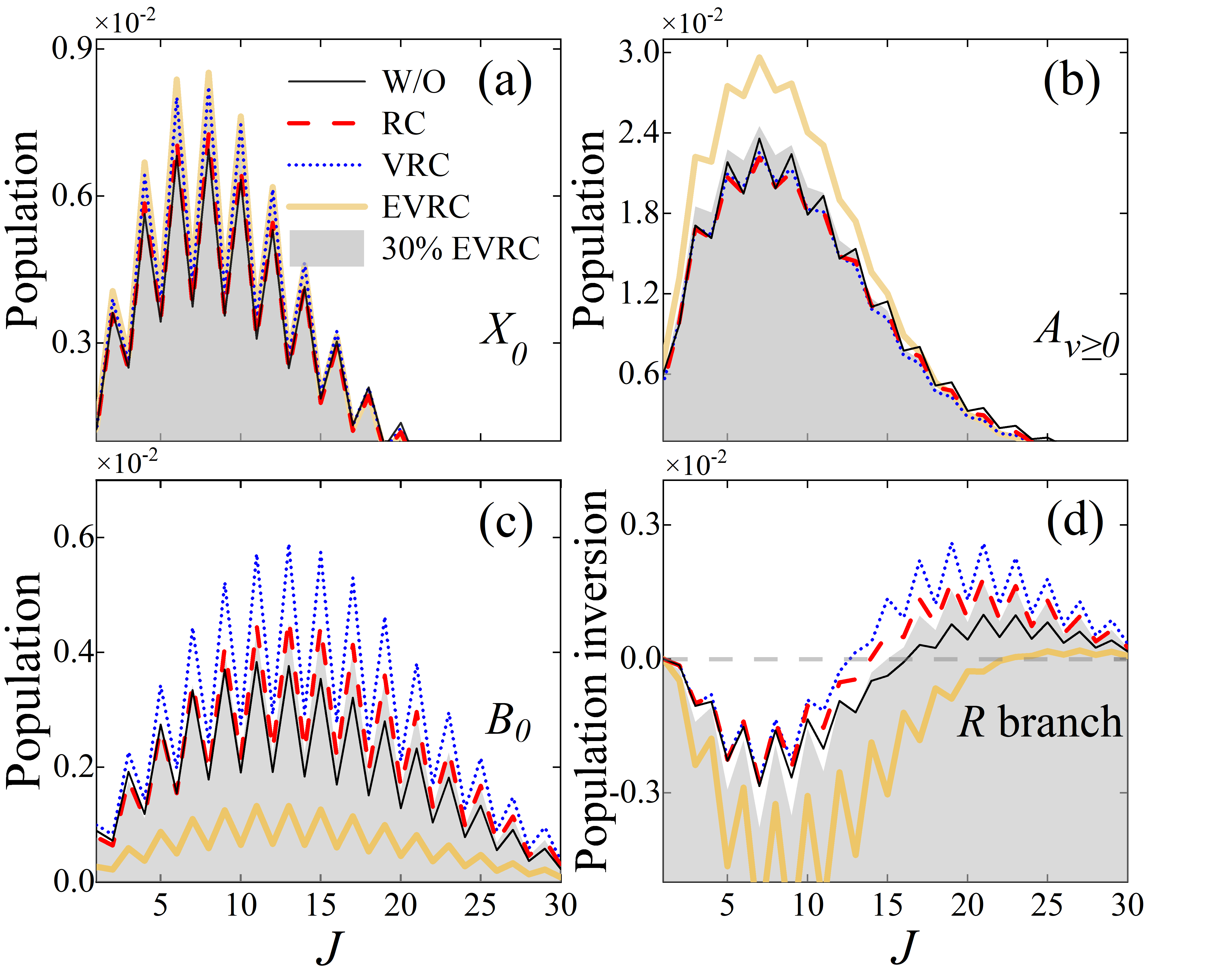}
\caption{(a)-(c) show the population of $X_{0J}$, $A_{v'\geq 0J}$ and $B_{0J}$ states as a function of $J$, respectively. (d) shows the $R$-branch population inversion between the $X_{0J-1}$ and $B_{0J}$ states. ``W/O" indicates results calculated without ionization-produced coherence. ``RC" include only the ionization-produced RCs. ``VRC" includes both the ionization-produced VCs and RCs. ``EVRC" includes ionization-produced ECs, VCs and RCs. ``30\%~EVRC" means the ECs are scaled to 30\% of their original values.}
\label{fig2}
\end{figure}

Figures~\ref{fig2}(a-c) show the populations of the rovibronic states $X_{0J}$, $A_{v'\ge 0J}$, and $B_{0J}$ obtained at the end of the pump pulse, considering different types of coherences. 
The populations for the same $J$ but different $M$ values have been summed.
In Fig.~\ref{fig2}(b), the populations of the $A_{v'\geq 0J}$ states are significantly higher compared to those of the $X_{0J}$ and $B_{0J}$ states. This is because the $X_0 \rightarrow A_{v'\geq 0}$ transition is nearly one-photon resonant under the 800-nm laser field, resulting in most of the $X_0$-state population being transferred to the $A_{v'\geq 0}$ state. Additionally, all vibrational quantum numbers are included for the $A_{v'\geq 0J}$-state populations.
In Fig.~\ref{fig2}(a), the $X_{0J}$-state populations peak at $J = 8$. In contrast, in Fig.~\ref{fig2}(c), the $B_{0J}$-state populations reach their peaks at higher $J$ values. These differences in the peak positions create population inversions between the $X_{0J+1}$ ($X_{0J-1}$) and $B_{0J}$ states at higher $J$ values, which will lead to lasing emission in the $P$ ($R$) branch. 
Figure~\ref{fig2}(d) shows only the population inversions for the $R$ branch. The $P$-branch results, which exhibit slightly higher population inversions, follow a similar trend to the $R$ branch and are therefore not shown here. 
 
\subsubsection{Rotational coherence}
First, we examine the effect of ionization-produced RC. 
By comparing the ``RC" results with the ``W/O" results in Figs.~\ref{fig2}(a-c), one can seen that the RCs substantially increase the population of the $B_{0J}$ state, while the populations of the $X_{0J}$ and $A_{v'\ge 0J}$ states remain largely unaffected. As a result, the population inversion between the $X_{0}$ and $B_{0}$ states is increased, as shown in Fig.~\ref{fig2}(d).

To understand the underlying mechanism, we employ a three-state $\Lambda$(or V-type) model for analysis. In this model, states 1 and 3 are nearly degenerate in energy, representing adjacent rotational levels within the same electronic state. They are optically coupled to state 2 via non-zero TDMs, which mimics the coupling to a rotational state of a different electronic state. The quantum Liouville equations for this three-state system are given as follows,
\begin{small}
\begin{equation}\label{eq:B1}
\begin{aligned}
\dot{\rho}_{22}&=-2u_{12}F(t)\text{Im}[\rho_{12}] -2u_{32}F(t)\text{Im}[\rho_{32}],  \\
\dot{\rho}_{12}&=-\text{i}\omega_{12}{\rho}_{12}+ \text{i}u_{12}F(t)(\rho_{22}-\rho_{11}) -\text{i}u_{23}F(t)\rho_{13},
\\
\dot{\rho}_{32}&=-\text{i}\omega_{32}{\rho}_{32}+ \text{i}u_{32}F(t)(\rho_{22}-\rho_{33}) -\text{i}u_{12}F(t)\rho_{13}^*,\\
....
\end{aligned}
\end{equation}
\end{small}%
If the instantaneous ionization-produced coherence $\rho_{13}^{\text{ins}}$ is inserted at $t - \delta t$, the contribution of this term to $\dot{\rho_{22}}(t)$ should be
\begin{equation}\label{eq:B2}
\dot{\rho}_{22}^{\text{ins}}(t)\approx 4u_{32}u_{12}F^2(t)\delta t \text{Re}[\rho_{13}^{\text{ins}}(t-\delta t)].
\end{equation}%
The sign of $u_{32}u_{12}\times \text{Re}[\rho_{13}^{\text{ins}}]$ determines whether $\rho_{13}^{\text{ins}}$ causes an increase or decrease in $\rho_{22}$. This effect essentially arises from a constructive or destructive interference between the $1\rightarrow 2$ and $3\rightarrow 2$ pathways.


\begin{figure} [htb]
\includegraphics[width=6.5cm,clip=true]{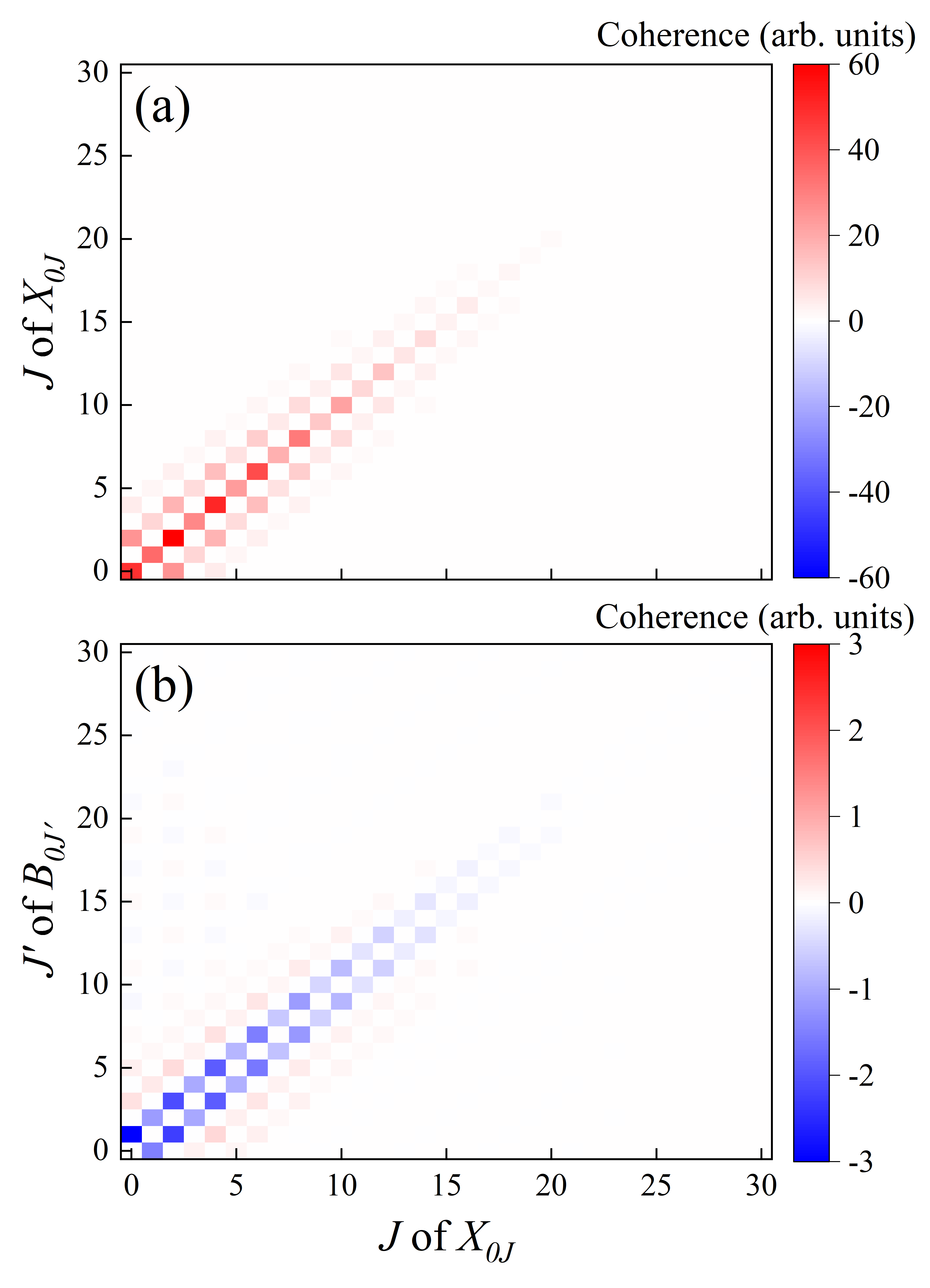}
\caption{(a) Rotational coherences between the $X_{0JM}$ and $X_{0J'M}$ states with $M=0$. Similar trends are observed for $M \ne 0$. (b) Electronic coherences between the $X_{0JM}$ and $B_{0J'M}$ states with $M=0$. Both are calculated with a field strength of 0.1~a.u.}
\label{fig3}
\end{figure}
  
For the $B_{0J}$ state, its population mainly comes from the $X_{vJ-1M}\rightarrow B_{0JM}$ and $X_{v'J+1M}\rightarrow B_{0JM}$ transitions (see Eq.~(\ref{eq:oij3})).
The $X_{v,J-1M}$, $B_{0,JM}$, and $X_{v',J+1M}$ states correspond to states 1, 2 and 3 in Eq.~(\ref{eq:B1}), respectively.
The instantaneous ionization-produced RC $\rho^{XX(\text{ins})}_{vJ-1M,vJ+1M}$ corresponds to $\rho_{13}^{\text{ins}}$.
According to Eq.~(\ref{eq:B2}), its influence is governed by its own sign and that of $O^{XB}_{vJ-1M,0JM}\times O^{XB}_{vJ+1M,0JM}$. 
From Eqs.~(\ref{eq:oij3}), both $O^{XB}_{vJ-1M,0JM}$ and $O^{XB}_{vJ+1M,0JM}$ share the same sign, so their product is positive. 
For the instantaneous ionization-produced RCs $\rho^{XX(\text{ins})}_{vJ-1M,vJ+1M}$, they are real quantities according to Eq.~(\ref{eq:insr}). Their values are shown in Fig.~\ref{fig3}(a). 
The RCs primarily appear in the diagonal region when $\Delta J $ is even, and their values are positive. Careful examination confirms that this behavior is consistent across different vibrational states and $M$ values. 
As a result,  $O^{XB}_{vJ-1M,0JM}O^{XB}_{vJ+1M,0JM}\times\rho^{XX\text{ins}}_{vJ-1M,vJ+1M}>0$. This means that the $X_{vJ-1M}\rightarrow B_{0JM}$ and $X_{vJ+1M}\rightarrow B_{0JM}$ pathways interfere constructively. The ionization-produced RCs increase the $B_{0JM}$-state population.


For the $A_{v'>0}$ states, their populations mainly come from three transition pathways: $X_{0J-1M}\rightarrow A_{v'JM}$, $X_{0JM}\rightarrow A_{v'JM}$ and $X_{0J+1M}\rightarrow A_{v'JM}$ (see Eqs.~(\ref{eq:oij3}). In Fig.~\ref{fig3}(a), ionization-produced $X$-state RCs appear only for even $\Delta J $. 
Therefore, their effect on the $X_0 \rightarrow A_{v'>0}$ transitions primarily arises from the interference between the $X_{0J-1M} \rightarrow A_{v'JM}$ and $X_{0J+1M} \rightarrow A_{v'JM}$ pathways. According to Eq.~(\ref{eq:B1}), this interference depends on the sign of the product $O^{XA}_{vJ-1M,v'JM}O^{XA}_{vJ-1M,v'JM}\times \rho^{XX(\text{ins})}_{vJ-1M,vJ+1M}$. Eqs.~(\ref{eq:oij3}) indicate that $O^{XA}_{vJ-1M,v'JM}$ and $O^{XA}_{vJ-1M,v'JM}$ have opposite signs, and Fig.~\ref{fig3}(a) shows that the RC $\rho^{XX(\text{ins})}_{vJ-1M,vJ+1M}>0$ is positive. As a result, their product is negative, meaning that the ionization-produced RCs suppress the $X_0 \rightarrow A_{v'>0}$ transition. This suppression explains why the ``RC" result in Fig.~\ref{fig2}(b) is slightly lower than the ``W/O" result. 

For the $X_0$ state, the combined effects of the enhanced $X_0\rightarrow B_0$ transition and the suppressed $X_0 \rightarrow A_{v'>0}$ transition result in a nearly unchanged $X_0$-state population. 
Ultimately, the ionization-produced RCs enhance the population inversion between the $X_{0}$ and $B_{0}$ states. Specifically, the population inversion between $X_{0,J}$ and $B_{0,J+1}$ is improved by approximately 80\%, as indicated by the red dashed line in Fig.~\ref{fig2}(d). 

To note, the ionization-produced RCs originates from the geometric alignment of N$_{2}^{+}$ \cite{posthumus1998dynamic}. These coherences influence the ionic populations by modulating the interference between rovibronic-state transition pathways. This effect is physically distinct from the phenomena observed in Fig.~\ref{fig1}(d), which is caused by the free evolution of the rotational wavepacket. In fact, these two effects jointly determine the overall influence of molecular rotation on the post-ionization dynamics of N$_{2}^{+}$ in the pump stage.

\subsubsection{Vibrational coherence}
Next, we examine the effect of the ionization-produced VCs.  
Comparing the ``VRC" results with the ``RC" results in Figs.~\ref{fig2}(a-c), it can be observed that the VCs significant increase the population of the $B_{0J}$ state, while the populations of the $X_{0J}$ and $A_{v'\ge 0J}$ states remain nearly unchanged at higher $J$. As a result, the population inversion between the $X_{0J}$ and $B_{0J-1}$ states is markedly enhanced due to the incorporation of the VCs, increasing by approximately 60\% compared to the ``RC" results as shown in Fig.~\ref{fig2}(d). These findings are consistent with the conclusions drawn from the vibronic model in our recent work \cite{xue2025}. The increase of the $B_{0}$ state population primarily results from constructive inteference between the $X_0\rightarrow B_0$ and $X_1\rightarrow B_0$ transition pathways. The underlying mechanism can also be interpreted using the three-state model described by Eq.~(\ref{eq:B1}). For a detailed discussion, please refer to Ref.~\cite{xue2025}.

\subsubsection{Electronic coherence}
Finally, let us examine the effect of the ionization-produced ECs, which can be evaluated by comparing the ``EVRC" results with the ``VRC" ones. As stated in Sec.~II~B, Eq.~(\ref{eq:insv}) predict the maximum DOC between electronic states, and so does its extension Eq.~(\ref{eq:insr}). In contrast, more accurate coherence models predict significantly lower electronic-state DOCs. Therefore, two scenarios of ECs are considered in our simulations. In scenario 1, the ECs are directly calculated by Eq.~(\ref{eq:insr}) with the degrees of ECs being 100\%, labeled as ``EVRC". In scenario 2, the ECs are reduced to 30\% of their original values, labelled as ``30\%~EVRC". 30\% is an estimated parameter used to mimic the ionization-produced ECs of N$_2^+$ predicted by the adiabatic strong-field approximation coherence model \cite{xue2025}. 

In scenario 1, we compare the ``EVRC" results with the ``VRC" results in Figs.~\ref{fig2}(a-c). As shown, the $X_0$-state population remains nearly unchanged, while the population of the $A_{v'\ge 0}$ states increase markedly. Meanwhile, the $B_0$-state population decreases significantly. As a result, the population inversion in the $R$ branch is significantly reduced, as shown in Fig.~\ref{fig2}(d). In fact, no inversion is achieved, which contradicts experimental observations, as strong $R$-branch 391-nm lasing signals are clearly detected under the current laser conditions \cite{zhang2021optical}. 
In scenario 2, we compare the ``30\%~EVRC" results with the ``VRC" results. Similar to the phenomena observed in scenario 1, the $X_0$-state population remains almost unchanged. The populations of the $A_{v'\ge 0}$ states increase slightly. While the $B_0$-state population decreases by only ~25\%, which is less pronounced than that in scenario 1.  
In this case, the ionization-produced ECs reduce the population inversion between the $X_0$ and $B_0$ states, but the $R$-branch population inversion remains achievable. 

The underlying mechanism can be revealed using a two-state model. For the two states interacting with an electric field $F(t)=  F_0  \text{cos}(\omega_0 t+\phi )$, the DM equations under the rotating wave approximation are given by
\begin{equation} \label{eq:2state1}
\begin{aligned}
&\dot{\tilde {\rho}}_{12}=\text{i}u_{12}\frac{F_0}{2} e^{-\text{i}\epsilon t+\text{i}\phi }
\Delta+\rho_{12}^{\text{ins}}e^{-\text{i}\omega_{21}t},\\
&\dot{\Delta}=-2u_{12}F_0\text{Im}\left ( \tilde {\rho}_{12}e^{\text{i}\epsilon t-\text{i}\phi } \right ) , 
\end{aligned}
\end{equation}%
where $\Delta=\rho_{22}-\rho_{11}$, $\rho_{12}=\tilde{\rho}_{12}e^{-\text{i}\omega_{12}t}$ and $\epsilon =\omega_{21}-\omega_0$. 
$\rho_{12}^{\text{ins}}$ represents the instantaneous ionization-produced EC.
Because $\tilde {\rho}_{12}(t)=\tilde {\rho}_{12}(t-\delta t)+\rho_{12}^{\text{ins}}(t-\delta t) e^{-\text{i}\omega_{21}(t-\delta t)}\delta t+O(\Delta)$, the population transfer caused by $\rho_{12}^{\text{ins}}$ can be assessed by
\begin{equation} \label{eq:2state2} 
\Delta^{\text{EC}} =-2u_{12}F_0\text{sin}(\epsilon \delta t)\!\!\int\!\!\text{Re}[\rho_{12}^{\text{ins}}(t)]   \text{cos}(\omega_0 t+\phi )  dt.  
\end{equation}%
In the derivation, the ECs are treated as real quantities, which is true according to Eqs.~(\ref{eq:insr}). 
Clearly, the sign of $\Delta^{\text{EC}}$ depends on the signs of $u_{12}$, $\epsilon$ and the integral. 
Taking the $X_{0J\pm 1}\rightarrow B_{0J}$ transition as an example, we analyse the influence of $\rho^{XB(\text{ins})}_{0J\pm 1,0J}$ on the transition dynamics, where $X_{0J\pm 1}$ and $B_{0J}$ corresponds to states 1 and 2, respectively. $u^{XB}_{0J\pm 1,0J}$ corresponds to $u_{12}$.
According to Eq.~(\ref{eq:oij3}), one has $\text{sgn}(u^{XB}_{0J\pm 1,0J})=\text{sgn}(u^{XB}_{00})$. Since $u^{XB}_{00}\approx -0.6~\text{a.u.} <0$, it follows that  $u^{XB}_{0J\pm 1,0J}<0$. Regarding $\epsilon$, at a laser wavelength of 800~nm, the detuning $\omega^{BX}_{0J,0J\pm 1}-\omega_0$ is positive. 
To evaluate the sign of the integral in Eq.~(\ref{eq:2state2}), we examine the instantaneous ionization-produced ECs under a static field of 0.09 a.u., as shown in Fig.~\ref{fig3}(b). 
The ECs $\rho^{XB(\text{ins})}_{0J\pm 1,0J}$ are negative along the diagonal. Further examination reveals that their signs reverse when the field is reversed. Accordingly, the integral is positive. 
As a result, $\Delta^{\text{EC}}<0$, indicating that ECs suppresses the $X_{0J\pm 1} \rightarrow B_{0J}$ transition. This explains the reduction in the $B_{0J}$-state population and, consequently, the decrease in population inversion when ionization-produced ECs are included. Similar analyses can be applied to the $X_{0J}\rightarrow A_{vJ'}$ transition dynamics. For a detailed discussion, please refer to Ref.~\cite{xue2025}.
 
In general, ionization-produced ionic coherences ultimately result in a substantial enhancement of the $X_{0J\pm 1}-B_{0J}$ population inversion and should not be neglected in the study of N$_2^+$ lasing. When all coherences are included (with ECs reduced to 30\% in accordance with the prediction in Ref.~\cite{xue2025}), the population inversion is approximately 80\% higher than in the ``W/O" case.   
  
\subsection{Seed propagation}
In this section, we investigate the propagation of the seed pulse through the N$_2^+$ gain medium. Prior to the arrival of the seed pulse, N$_2^+$ has been prepared by the pump pulse. We utilize the ionic DM generated under the ``30\% EVRC" scenario as the initial condition for the seed propagation, with the corresponding rovibronic-state populations shown as the grey areas in Fig.~\ref{fig2}.
The time-delayed seed pulse is modeled as a 400-nm, 100-fs Gaussian-shaped, linearly polarized laser pulse with a peak intensity of $10^{9}\text{W/cm}^2$. The gas pressure is set to 30 mbar. 
The seed propagation is simulated by solving Eqs.~(\ref{eq:MAXWELL}) and (\ref{eq:BLOCH}). To analyze the absorption and emission characteristics of the seed pulse, we define a difference spectrum as follows
\begin{equation}\label{eq:dlasing}
   S^{\text{Diff}}(\omega)=|F^{\text{lasing}}(\omega)| ^ 2 -|F^{\text{seed}}(\omega)| ^ 2,
\end{equation}
where $|F^{\text{seed}}(\omega)|^2$ and $|F^{\text{lasing}}(\omega)|^2$ represent the spectra of the input seed and output lasing signals, respectively.

\begin{figure} [htb]
\includegraphics[width=9cm,clip=true]{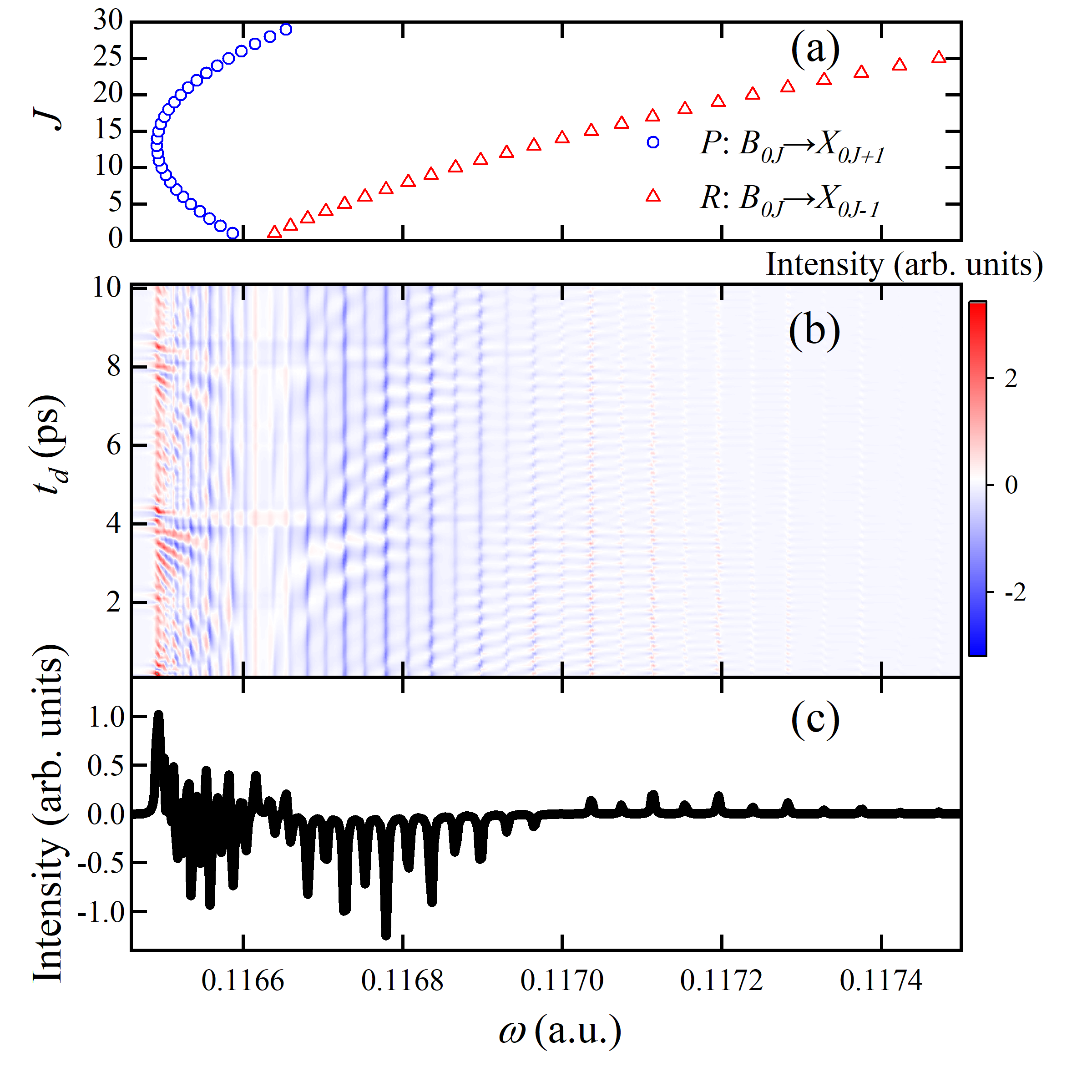}
\caption{(a) $P$- and $R$-branch frequencies as a function of $J$. (b) Lasing spectra as functions of frequency and seed delay time $t_d$, calculated for a propagation length of 0.1~mm. (c) Delay-averaged lasing spectrum.}
\label{fig4}
\end{figure}

\subsubsection{Without considering macroscopic propagation}
We first examine the response of the seed pulse on the N$_2^+$ ensemble, with macroscopic propagation effects excluded. For this purpose, the propagation distance is set to a very small value of 0.1~mm. Figure~\ref{fig4}(b) shows the difference spectrum as a function of the delay time $t_{d}$.
The spectrum can be divided into the $P$ and $R$ branches.
Corresponding frequencies as a function of $J$ are depicted in Fig~\ref{fig4}(a).  
In Fig.~\ref{fig4}(b), the spectrum exhibits oscillatory absorption and gain with sub-picosecond periods along $t_d$. Similar features have also been observed both experimentally and theoretically \cite{arissian2018transient,richter2020rotational}. 
These oscillations originate from interferences between different rovibronic transition pathways. For the $X_{0,J\pm 1} \rightarrow B_{0,J}$ absorption pathways, constructive (or destructive) interference occurs when the phase of $\rho^{XX}_{0J-1,0J+1}$ satisfies $\phi^{XX}_{0J-1,0J+1}=2N\pi$~[or $(2N+1)\pi$], where $N$ is an integer. Under this condition, absorption at the frequency $\omega^{BX}_{0J,0J\pm 1}$ is enhanced (or suppressed). Similarly, for the $X_{0,J} \leftarrow B_{0,J\pm 1}$ emission pathways, constructive (or destructive) interference occurs when $\phi^{BB}_{0J-1,0J+1}=2N\pi$ [or $(2N+1)\pi$], leading to enhanced (or suppressed) emission at $\omega^{BX}_{0J\pm 1,0J}$. 
Taking the $R$ branch as an example, the $J$th signal at $\omega=\omega^{BX}_{0J,0J-1}$ will exhibit delay-dependent oscillations with frequencies $\omega^{XX}_{J-1,J+1}$, $\omega^{XX}_{J-1,J-3}$, and $\omega^{BB}_{J,J\pm 2}$. According to Eq.~(\ref{energy}), these frequencies increase with $J$. Therefore, the oscillations in the $R$ branch become faster as $J$ (and thus $\omega$) increases, as shown in Fig.~\ref{fig4}(b).

Figure~\ref{fig4}(c) shows the delay-averaged difference spectrum over a 10-ps delay range, in which both absorption and emission lines are clearly visible. Since the $P$-branch signal arises from multiple emission pathways, whereas each order of the $R$-branch signal is contributed by a single pathway (see Fig.~\ref{fig4}(a)), we focus our discussion on the $R$-branch for clarity. In the $R$ branch, absorption occurs for $J < 15$, while emission takes place for $J\ge 15$. This behavior is consistent with the population inversion under the ``30\%~EVRC" condition, as shown by the grey area in Fig.~\ref{fig2}(d). These results demonstrate that, in the absence of macroscopic propagation, the output lasing spectrum directly reflects the population inversion between the $B_{0J-1}$ and $X_{0J}$ states in the N$_2^+$ ensemble.

\subsubsection{With consideration of macroscopic propagation}
\begin{figure} [htb]
\includegraphics[width=9cm,clip=true]{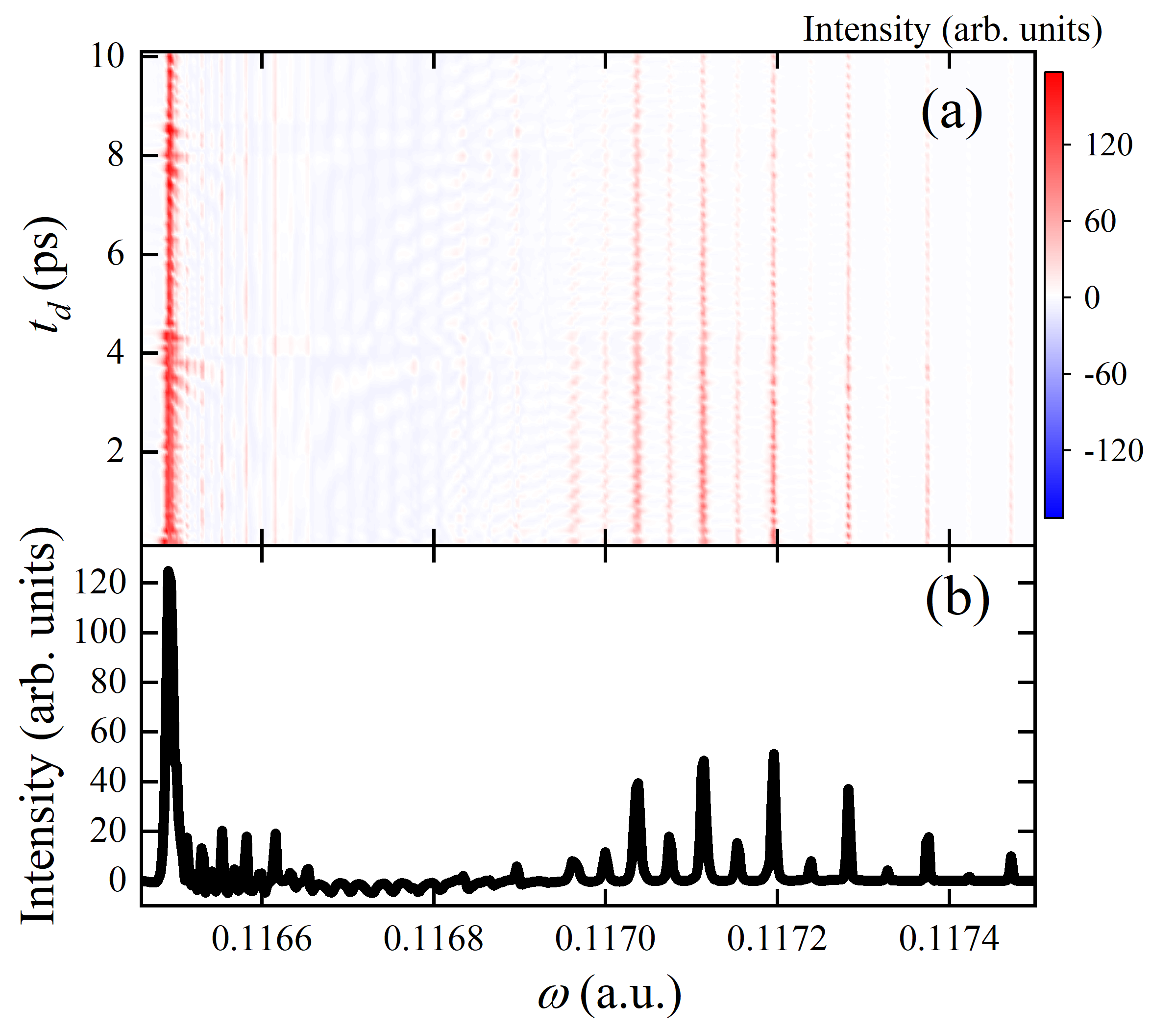}
\caption{(a) Lasing spectra as functions of frequency and seed delay time, calculated for a propagation length of 2.5~mm. (b) Delay-averaged lasing spectrum.}
\label{fig5}
\end{figure}

We now examine the lasing signal with macroscopic propagation effects included. The propagation length is set to $z_{\text{max}}=2.5$~mm. Figure~\ref{fig5}(a) shows the corresponding difference spectrum as a function of $t_{d}$. The oscillatory features along $t_d$ are still visible, originating from the RCs within the $X_0$ and $B_0$ states. Figure~\ref{fig5}(b) presents the delay-averaged spectrum. In contrast to Fig.~\ref{fig4}(c), the absorption lines nearly vanish, and the overall emission intensity is significantly enhanced. In the $R$ branch, seed amplification is even observed for the $X_{0J-1}-B_{0J}$ state pair, whose population inversion is not established (see the ``30\% EVRC" result in Fig.~(\ref{fig2})(d)). This observation clearly demonstrates that population inversion is not a necessary condition for lasing generation. We suggest that RCs may play a significant role in lasing generation upon macroscopic propagation.

\begin{figure} [htb]
\includegraphics[width=8cm,clip=true]{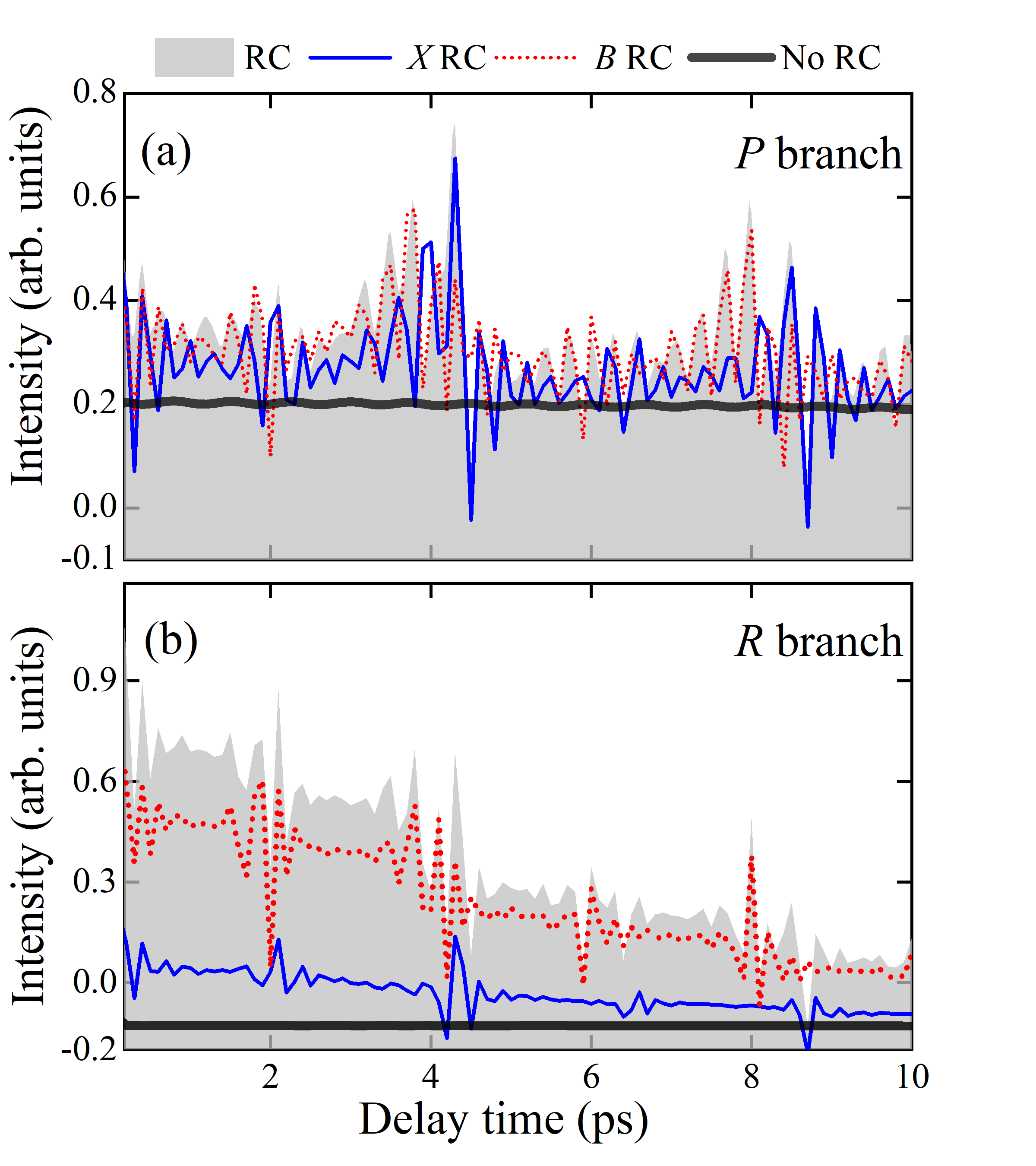}
\caption{Integrated lasing signal intensities as a function of seed delay time $t_d$: (a) $P$ branch and (b) $R$ branch. Labels are defined in the main text.}
\label{fig6}
\end{figure}

To evaluate the effects of RCs, we perform additional calculations under three artificial RC scenarios: (1) considering only the $X$-state RCs (denoted as ``$X$ RC"), (2) considering only the $B$-state RCs (denoted as ``$B$ RC"), and (3) excluding both the $X$- and $B$-state RCs (denoted as ``No RC"). Additionally, the calculation that include all RCs is denoted as ``RC". 
Figures~\ref{fig6}(a) and \ref{fig6}(b) show the delay-dependent lasing signals integrated over the $P$- and $R$-branches, respectively.  
As shown, the ``RC" results produce the strongest signals in both branches. The ``$B$ RC" result closely overlaps with the ``RC" result in the $P$ branch, but is slightly weaker in the $R$ branch. The ``$X$ RC" results are noticeably weaker in both branches. Finally, the ``No RC" results give the weakest signals, which remain nearly constant with $t_d$ due to the lack of RCs. 
Notably, in Fig.~\ref{fig6}(b), the ``No RC" result is even negative, indicating a net absorption of the seed pulse in the $R$ branch. This can be understood by referring to the ``30\% RVEC" results Fig.~\ref{fig2}(d), where the area of the non-inversion region is slightly larger than that of the inversion region relative to the zero horizon line. 
In addition, the signals generally decrease as $t_d$ increases, especially for the $R$ branch. This phenomenon, commonly observed in experiments \cite{zhang2013rotational,xie2014coupling}, can be attributed to the decoherence effects governed by the dephasing time $T_d$ in Eq.~(\ref{eq:BLOCH}).    
These comparisons across different RC scenarios demonstrate that RCs are crucial for seed amplification and can significantly enhance the lasing signal.

\begin{figure} [htb]
\includegraphics[width=9cm,clip=true]{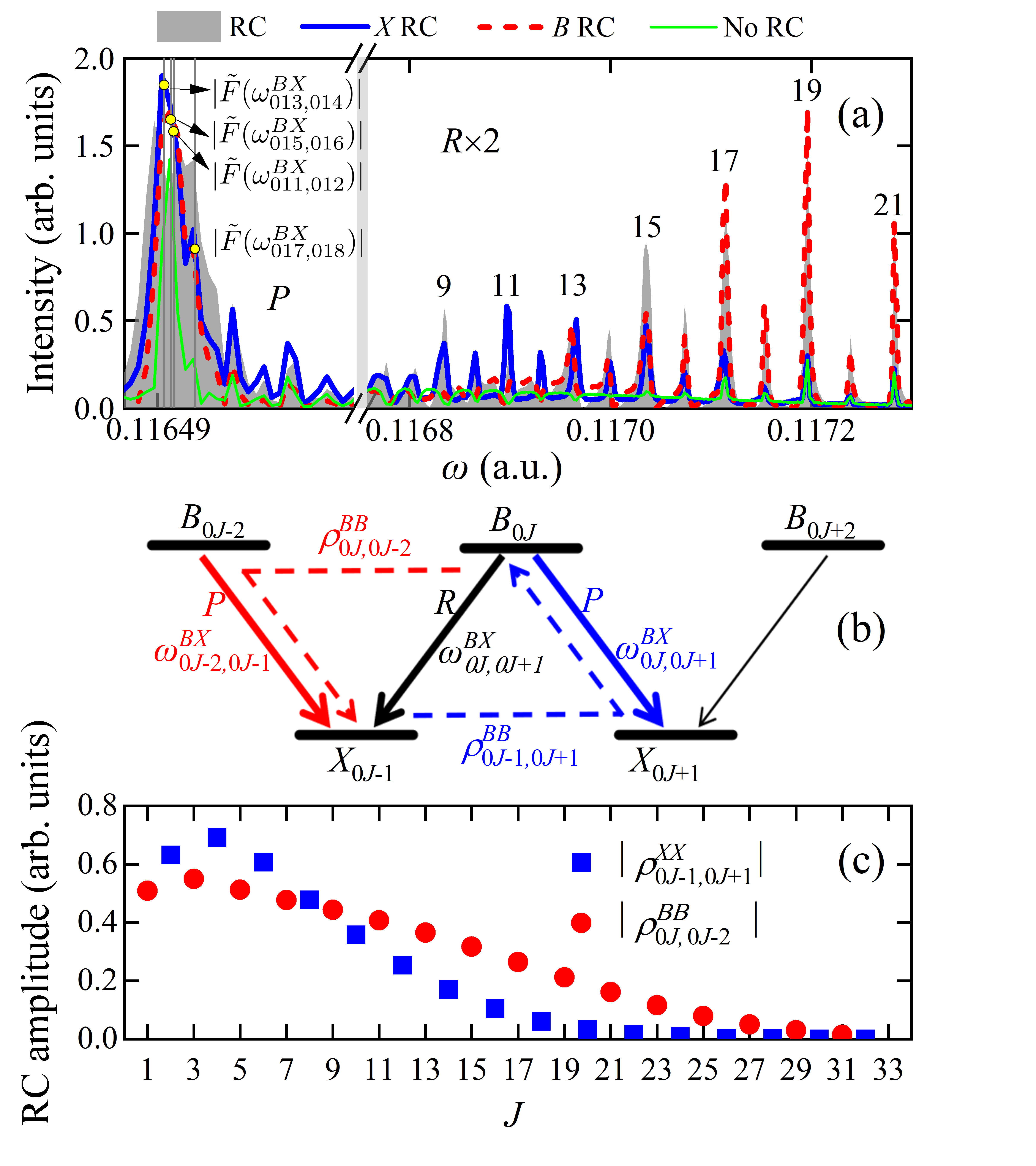}
\caption{(a) Lasing spectra at a seed delay time of $t_{d} = 4.3$~ps. Labels are defined in the main text. $P$-branch frequency as a function of $J$ are shown as a purple solid line for reference (right axis). (b) Schematic representation of the emission mechanism for the N$_2^+$ 391-nm lasing. (c) Amplitudes of the rotational coherences in the $X$ and $B$ electronic states with $M =0$.}
\label{fig7}
\end{figure}

Figure~\ref{fig7}(a) shows the lasing spectra at a fixed delay time of $t_{d}=4.3$~ps, where the signal is relatively stronger (see Fig.~\ref{fig6}). As shown, when RCs are excluded (``No RC"), lasing signals are generated at orders where population inversions are established (see the grey area in Fig.~\ref{fig2}(d)). But the signals are extremely weak. In contrast, when RCs are included, the ``RC" signals are significantly stronger than the ``No RC" signals for both the $P$- and $R$-branches. These results demonstrate that, although both population inversion and RCs can contribute to lasing, the latter plays the dominant role. 
For simplicity, we focus on the $R$ branch to analyze the role of RCs in seed amplification. 
Comparing the ``$X$ RC" spectrum with the ``RC" spectrum, the $R$-branch signals show a noticeable weakening in the population inversion region ($J> 15$; see the grey area in Fig.~\ref{fig2}(d)), while exhibiting little change in the non-inversion region ($J < 15$). However, when comparing the ``$B$ RC" spectrum with the ``RC" spectrum, the $R$-branch signals remain almost unchanged in the inversion region ($J> 15$), but are significantly reduced in the non-inversion region ($J < 15$). 
These results suggest that the $X$-state RCs primarily enhance the signals in the non-inversion region, while the $B$-state RCs play a dominant role in the inversion region.

The mechanism of seed amplification can be analyzed using the $\Lambda$ (or V)-type three-state model. 
The three-state structure is described in Sec.~III B1.
The coherence between states 1 and 3, $\rho_{13}=\tilde{\rho}_{13}e^{-i\omega_{13}t}$, represents the RC. For the propagation of a seed electric field $F(t,Y)=\frac{1}{2} \tilde{F}(t,Y)e^{-i\omega_s t}+c.c.$, the simplified MB equations in the retarded frame can be expressed as
\begin{equation}\label{eq:C1}
\begin{aligned}
\frac{\partial }{\partial z} \tilde{F}  ( t,Y)&= \text{i}\alpha _{12}\tilde{\rho}_{21}e^{-\text{i}(\omega_{21}-\omega_s)t}+ \text{i}\alpha _{32}\tilde{\rho}_{23}e^{-\text{i}(\omega_{23}-\omega_s)t}\\
\tilde{\rho}_{12}&=\text{i}F( t,Y  )e^{-\text{i}\omega_{21}t}(d_{12}\Delta_{21}-\rho_{13}d_{32} )\\
\tilde{\rho}_{32}&=\text{i}F( t,Y  )e^{-\text{i}\omega_{23}t}(d_{32}\Delta_{23}-\rho_{31}d_{12} ),
\end{aligned}
\end{equation}
where $\alpha _{ij}$ denotes the propagation parameter defined in Eq.~(\ref{eq:MAXWELL}). $\Delta_{ij} = \rho_{ii} - \rho_{jj}$ is the population difference. Then the electric field propagation can be expressed as
\begin{equation}\label{eq:C2}
\begin{aligned}
&\frac{\partial }{\partial Y} \tilde{F} ( t,Y  )\\
&\approx\alpha _{21}e^{-i(\omega _{21}-\omega_s)t} \left [ u_{12} \Delta  _{21}\tilde{F} ( \omega _{21},Y ) - u_{32}\tilde {\rho}_{31}  \tilde{F}^* ( \omega _{23},Y )  \right ]\\
&+\alpha _{23}e^{-i(\omega _{23}-\omega_s)t} \left [ u_{32} \Delta _{23}\tilde{F} ( \omega _{23},Y )- u_{12}\tilde {\rho}_{13}  \tilde{F}^* ( \omega _{21},Y )  \right ],
\end{aligned}
\end{equation}
where $\tilde{F} (\omega_{ij},Y)$ is the spectral amplitude at frequency $\omega_{ij}$ and position $Y$. In the derivation, $\Delta_{ij}$ and the slow-varying RC $\tilde{\rho}_{13}$ are treated as constants.
This equation indicates that the lasing amplification at frequencies $\omega_{21}$ and $\omega_{23}$ arises from two contributing components: the population differences $\Delta_{21}$ and $\Delta_{23}$, and the RC $\tilde{\rho}_{13}$. Their contributions are modulated by the accompanying spectral amplitudes $\tilde{F}(\omega_{ij},Y)$.

According to Eq.~(\ref{eq:C2}), the $R$-branch $J$th-order lasing signal is influenced by three transition pathways: $B_{0J}\rightarrow X_{0J-1}$, $B_{0J-2}\rightarrow X_{0J-1}$, and $B_{0J}\rightarrow X_{0J+1}$, as shown by the thick solid arrows in Fig.~\ref{fig7}(b). The contribution of the first pathway is determined by the population inversion $\rho^{BB}_{0J,0J}-\rho^{XX}_{0J-1,0J-1}$, modulated by $\tilde{F}(\omega^{BX}_{0J,0J-1})$. Physically, it means that the population inversion drives lasing amplification, directly stimulated by the optical signal at the resonant frequency. 
The last two pathways are contributed by RCs, governed by $\rho^{BB}_{0J,0J-2}\times\tilde{F}(\omega^{BX}_{0J-2,0J-1})$ and $\rho^{XX}_{0J-1,0J+1}\times\tilde{F}(\omega^{BX}_{J,J+1})$, respectively. These terms indicate that the RCs in the $B$ and $X$ state facilitate lasing amplification, indirectly stimulated by the optical signals at $\omega^{BX}_{0J-2,0J-1}$ and $\omega^{BX}_{0J,0J+1}$ in the $P$ branch, respectively. These interplays are visually highlighted by the red and blue dashed arrows in Fig.~\ref{fig7}(b). 
For the $R$-branch signals at $J > 15$, namely in the population inversion region, the stimulating signal associated with the $B$-state RCs, $|\tilde{F}(\omega^{BX}_{0J-2,0J-1})| $, is generally stronger than that associated with the $X$-state RCs, $|\tilde{F}(\omega^{BX}_{0J,0J+1})|$. For example, for the $R$-branch 17th signal, $|\tilde{F}(\omega^{BX}_{015,016})|>|\tilde{F}(\omega^{BX}_{017,018})| $, as shown by the dots on the red dashed line in Fig.\ref{fig7}(a). As a result, the $B$-state RCs play a more prominent role for the lasing generation.
In contrast, for the $R$-branch signal at $J<15$, namely in the non-inversion region, $|\tilde{F}(\omega^{BX}_{J-2,J-1})|$ is generally smaller than $|\tilde{F}(\omega^{BX}_{J,J+1})|$. For example, for the $R$-branch 13th signal, $|\tilde{F}(\omega^{BX}_{011,012})|<|\tilde{F}(\omega^{BX}_{013,014})| $, as shown by the dots on the blue solid line in Fig.\ref{fig7}(a). Therefore, the $X$-state RCs contribute more significantly.  
On the other hand, the contribution of RCs also depend on their absolute values. Figure \ref{fig7}(c) shows the absolute values of the ionic RCs produced by the pump pulse.    
In the population inversion region with $J> 15$, $|\rho^{BB}_{0J,0J-2}|$ is larger than $|\rho^{XX}_{0J-1,0J+1}|$. Conversely, for $J\le 8$ in the non-inversion region, $|\rho^{XX}_{0J-1,0J+1}|$ is larger. 
The above dual-perspective analysis confirms that the $B$-state RCs contribute more significantly in the population inversion region, whereas the $X$-state RCs play a dominant role in the non-inversion region.

\section{Conclusions}
In conclusion, we investigate the influence of molecular rotation on the generation of N$_{2}^{+}$ lasing at 391 nm. By using rovibronic states as the basis set and solving the open-system DM equations, we explore the formation of the N$_2^+$ ensemble induced by the femtosecond pump pulse. It is found that molecular rotation can noticeably change the angle-dependent populations of vibronic states in N$_2^+$ within just tens of femtoseconds, even though the rotational period is on the picosecond timescale. 
Furthermore, ionic coherences generated upon strong-field ionization will affect the transition dynamics within N$_2^+$. Specifically, the RCs will enhance the population inversion between the $X_0$ and $B_0$ states. When all rotational, vibrational, and electronic coherences are considered, the population inversion can increase by up to $\sim 80\%$. 
Finally, by solving the MB equations, we investigate the seed propagation process and the mechanism of lasing generation.
It is found that both the population inversion and RCs within the $X_0$ and $B_0$ states contribute to lasing amplification, with the latter playing a dominant role. In the presence of population inversion, RCs significantly enhance the lasing intensity.

This study highlights the important role of molecular rotation in N$_2^+$ lasing and provides a theoretical foundation for future experimental investigations of the rotational effects.

\section*{ACKNOWLEDGEMENT}
This work was supported by the National Natural Science Foundation of China (Grants No.~12274188, No.~12004147, No.~12204209), the Natural Science Foundation of Gansu Province (Grant No.~23JRRA1090), and the Fundamental Research Funds for the Central Universities (Grant No.~lzujbky-2023-ey08).

	\bibliography{name}

\end{document}